\documentstyle[twoside,fleqn,npb1,epsfig]{article}
\title{\normalsize{UB-HET-02-08\\
November 2002\\[3.mm]}\Large
Electroweak Radiative Corrections to Weak Boson Production at
Hadron Colliders}

\author{U.~Baur and D.~Wackeroth\address{Physics Department \\
State University of New York at Buffalo\\
       Buffalo, NY 14260, USA}%
}

\begin{document}

\begin{abstract}
We summarize the status of calculations of the electroweak radiative 
corrections to W and Z boson
production via the Drell-Yan mechanism at hadron colliders. To fully 
exploit the precision physics potential of the
high-luminosity environment of the Fermilab Tevatron $p \bar p$ (Run~II)
and the CERN 
LHC $pp$ colliders, it is crucial that the theoretical predictions are
well under control.  The envisioned precision physics program includes
a precise measurement of the W boson mass and width, and the (leptonic) weak
mixing angle, as well as probing the Standard Model (SM) of
electroweak interactions at the highest accessible center-of-mass
energies. Some numerical results are presented. 
\end{abstract}

\maketitle

\section{Introduction}

The Standard Model of electroweak interactions (SM) so far withstood
all experimental challenges and is tested as a quantum field theory at
the 0.1\% level~\cite{Abbaneo:2001ix}. However, the mechanism of mass
generation in the SM predicts the existence of a Higgs boson which, so
far, has eluded direct observation. Direct searches at LEP2 give 
a 95\% confidence-level lower bound on the mass of the
SM Higgs boson of $M_H> 114.4$~GeV~\cite{Grunewald:2002wg}.  Indirect
information on the mass of the Higgs boson can be extracted from the
$M_H$ dependence of radiative corrections to the W boson mass. With
the present knowledge of the W boson and top quark masses, and the
electromagnetic coupling constant, $\alpha(M_Z^2)$, the SM
Higgs boson mass can be indirectly constrained to 
$M_H=88^{+53}_{-35}$~GeV~\cite{Abbaneo:2001ix} by a global fit to all 
electroweak precision
data.  Future more precise measurements of the W boson and top quark
masses are expected to  considerably improve the present indirect bound on
$M_H$:~with a precision of 27~MeV (16~MeV) for the W boson mass, $M_W$, and
2.7~GeV (1.4~GeV) for the top quark mass which are target values for
Run~IIa (Run~IIb) of the Tevatron~\cite{Baur:2001yp}, 
$M_H$ can be predicted with an uncertainty
of about $35\%$ (25\%). In addition, the confrontation of a precisely measured
W boson mass with the indirect SM prediction from a global fit to all
electroweak precision data, $M_W=80.398 \pm 
0.019$~GeV~\cite{Abbaneo:2001ix}, together with a precise measurement of
the W width, $\Gamma_W$, from the high transverse mass tail, 
will provide a stringent test of
the SM. A detailed discussion of the prospects for the precision
measurement of $M_W$ and $\Gamma_W$, and of the (leptonic) effective
weak mixing 
angle, $\sin^2\theta_{eff}^l$, at Run~II and the
LHC is given in Refs.~\cite{Brock:1999ep} and~\cite{Haywood:1999qg},
respectively.

In order to measure $M_W$ with high precision in a hadron collider
environment it is necessary to fully control higher order QCD and 
electroweak radiative corrections to the W and Z production processes.
The status of the QCD corrections to W and Z boson production 
at hadron colliders is reviewed in Refs.~\cite{Catani:2000jh,Baur:2000xd}. 
Here we discuss the electroweak ${\cal O}(\alpha)$ corrections to 
$p\,p\hskip-7pt\hbox{$^{^{(\!-\!)}}$} \to W^{\pm} \to l^{\pm} \nu_l$ and 
$p\,p\hskip-7pt\hbox{$^{^{(\!-\!)}}$} \to \gamma^*, Z \to l^+ l^-$ 
($l=e,\mu$) as
presented in detail in Refs.~\cite{Wackeroth:1997hz,Baur:1999kt} 
and~\cite{Baur:1998wa,Baur:2001ze,Dittmaier:2001ay,Baurinprep},
concentrating on photon radiation effects and the impact of
non-resonant weak corrections on the W width measurement.

\section{Photon Radiation Effects}

We illustrate the effects caused by photon radiation using Z boson
production as an example. The electroweak ${\cal O}(\alpha)$ corrections
to neutral-current 
Drell-Yan processes naturally decompose into QED and weak
contributions, i.e.~they build gauge invariant subsets and thus can
be discussed separately. The radiation of collinear photons off quarks
requires the 
factorization of the arising mass singularities into the PDFs which
introduces a dependence on the QED factorization scale, $\mu_{QED}$. 

\begin{figure}[h!]
\begin{center}
\epsfig{file=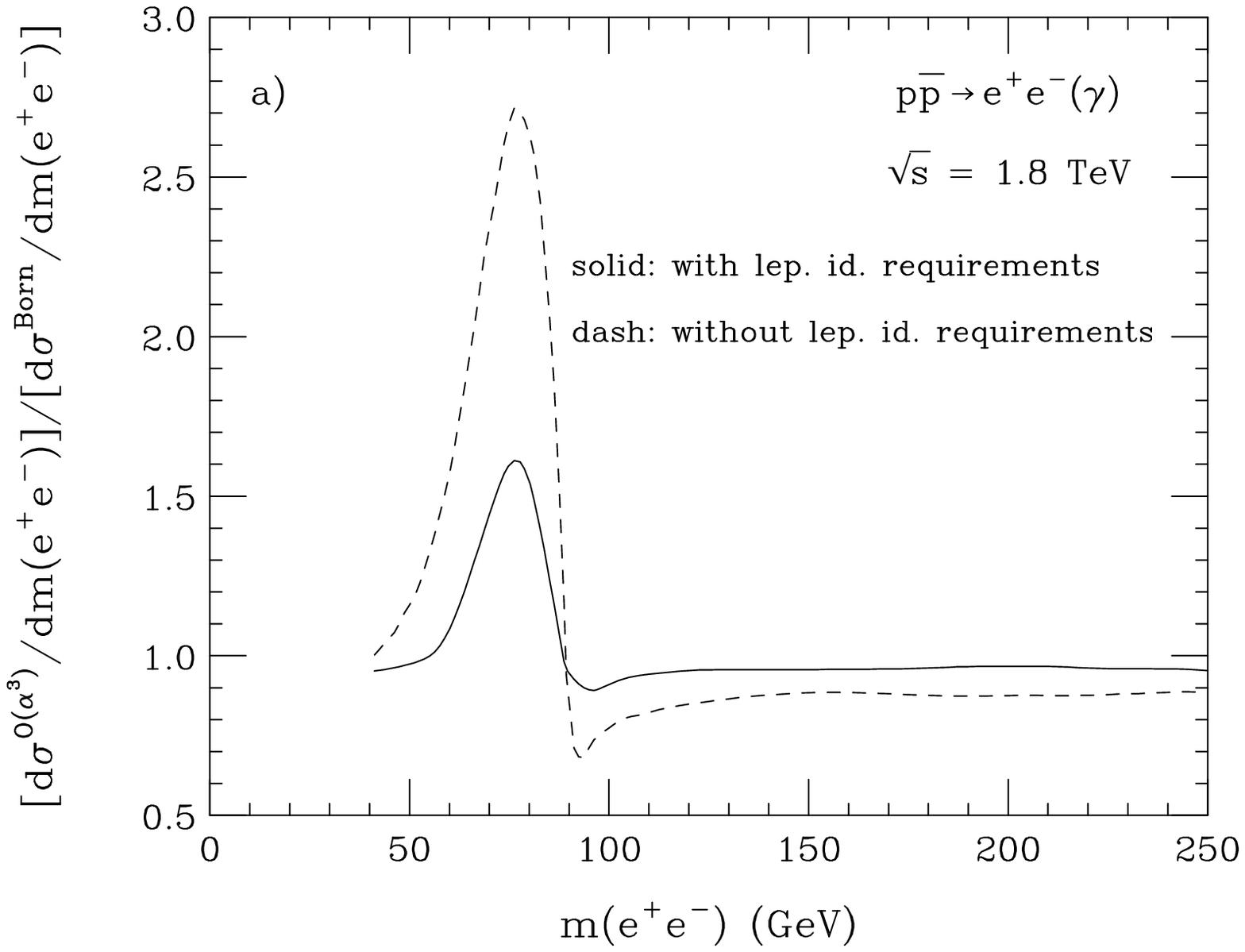,height=2.25in}\\
\epsfig{file=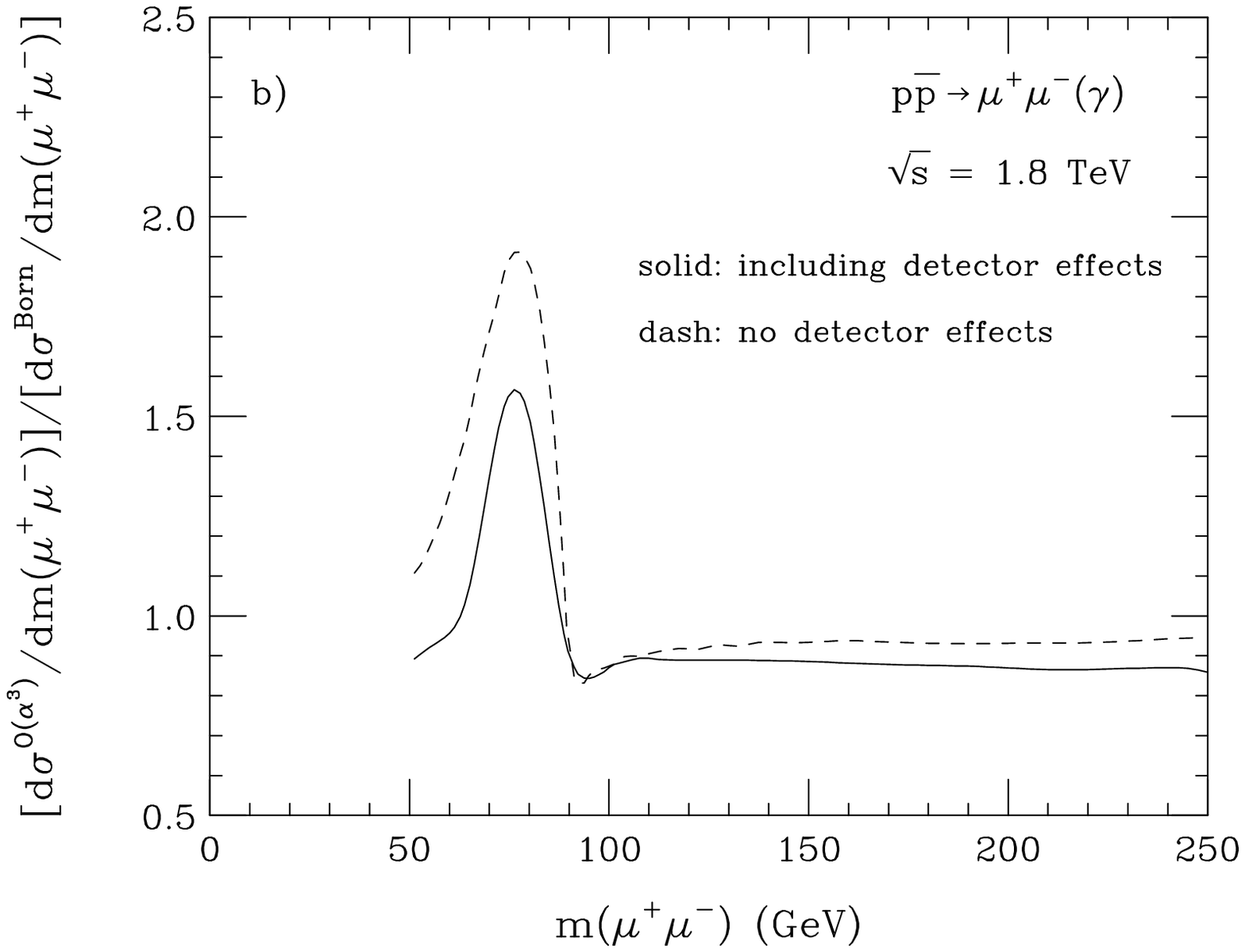,height=2.25in}
\caption[0]{The relative corrections to the $m(e^+ e^-)$ and
$m(\mu^+ \mu^-)$ distributions in Drell-Yan production at the Tevatron 
due to the ${\cal O}(\alpha)$ QED corrections 
(from Ref.~\cite{Baur:1998wa}). }\label{fig:one}
\end{center}
\end{figure}

In Fig.~\ref{fig:one} we show the effect of the QED corrections on
the invariant mass distribution of the final state lepton pair. Photon
radiation strongly affects the shape of the $m(l^+l^-)$
distribution. This is due to the mass singular terms associated with
final state photon radiation, which are proportional to $\log(\hat
s/m_l^2)$, where $m_l$ is the mass of the final state lepton.
At the Z pole, the differential cross section is reduced by about
30\% in the electron case and by about 20\% in the muon case. Below the
Z peak, QED corrections enhance the cross section by up to a factor~2.7
(1.9) for electrons (muons). 

When lepton identification requirements are taken into
account, the large contributions from mass singular logarithms
largely cancel in the electron case because electron and photon momenta
are combined for 
small opening angles between the two particles. For muons,
the energy of a photon which traverses the same calorimeter cell as the
muon is required to be $E_\gamma<2$~GeV, and only the momenta of muons and 
photons with $E_\gamma<2$~GeV are combined. As a result, the
mass singular logarithmic terms survive, and the effect of radiative
corrections becomes more pronounced, especially at large di-lepton
masses. 

The distortion of the shape of the Breit-Wigner resonance curve
of the Z boson due to final state photon radiation leads to a shift in
the Z boson mass extracted from data. In the electron channel, the shift
is about 100~MeV, whereas for $Z\to\mu^+\mu^-$ one finds approximately
300~MeV. 

\begin{figure}[h!]
\begin{center}
\epsfig{file=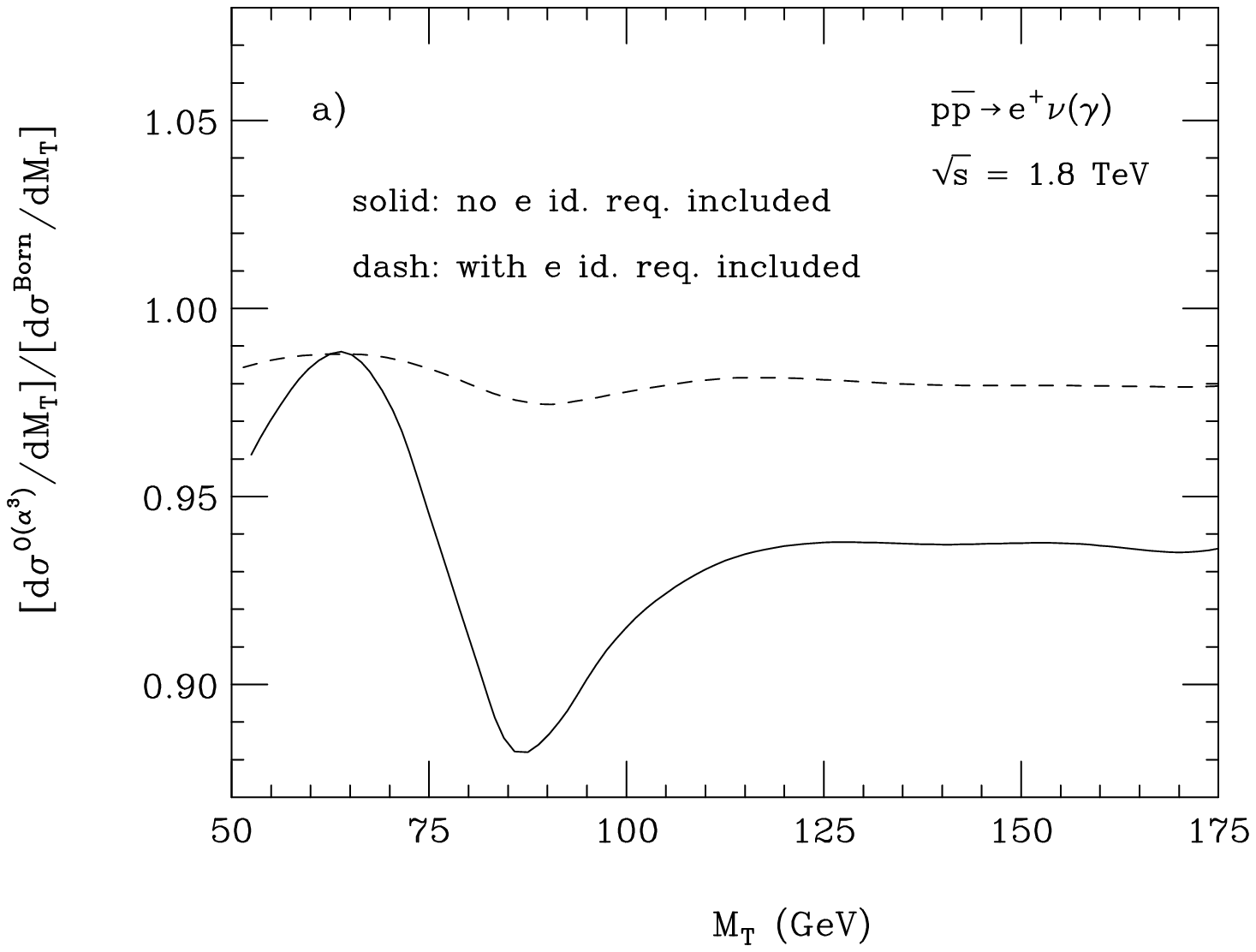,height=2.25in}\\
\epsfig{file=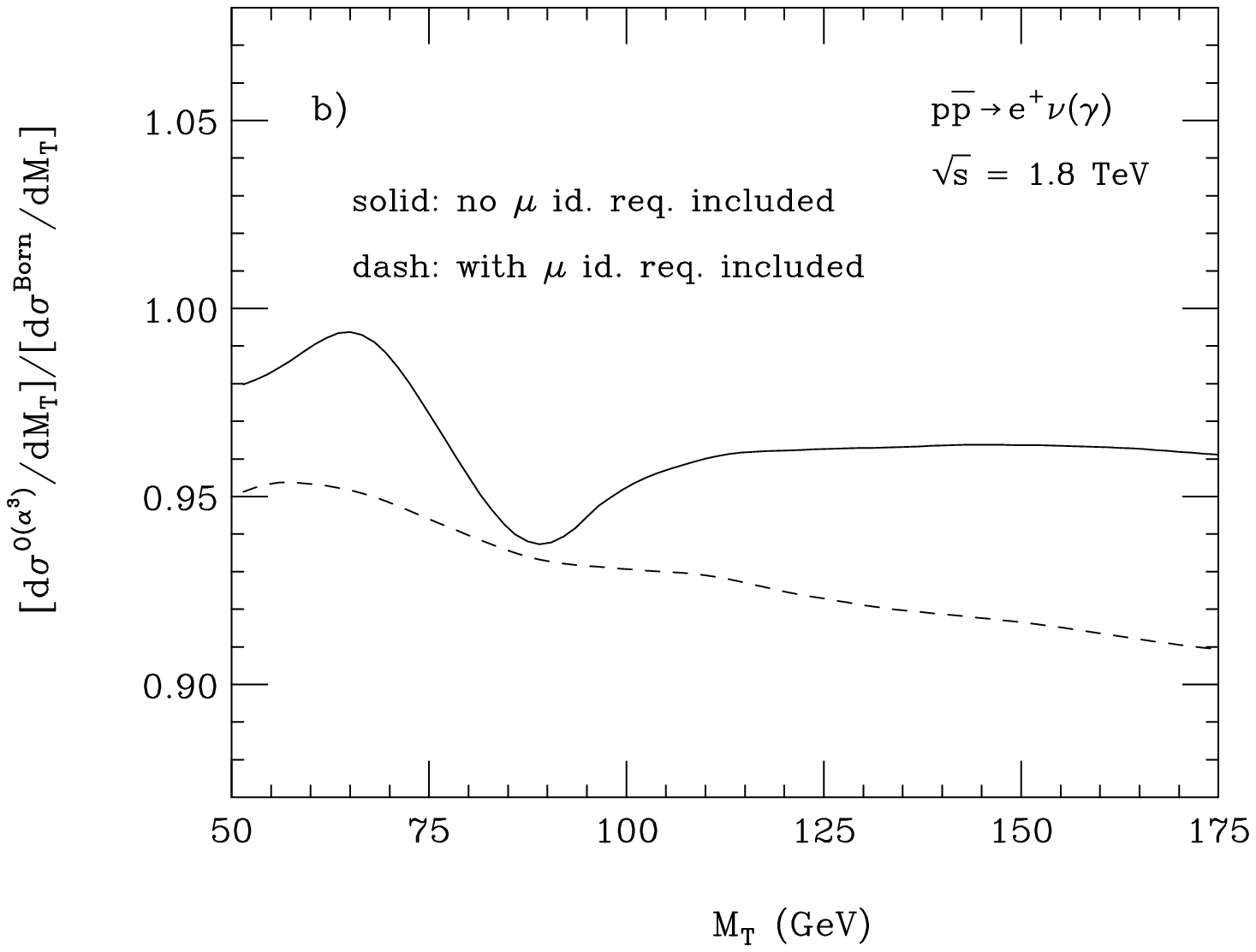,height=2.25in}
\caption[0]{The relative corrections to the $M_T(l \nu)$ distributions
at the Tevatron when taking into account electroweak
${\cal O}(\alpha)$ corrections
(from Ref.~\cite{Baur:1999kt}). }\label{fig:onea}
\end{center}
\end{figure}

Since the neutrino in W decays is not observed, it is impossible to
reconstruct the $l\nu_l$ invariant mass, and the transverse mass ($M_T$)
distribution is frequently used to extract the mass and width of the
W boson. Final state photon radiation in $W\to l\nu_l$ affects the shape of
the $M_T$ distribution in a similar way as the di-lepton invariant mass
distribution in $Z\to l^+ l^-$. The ratio of
the ${\cal O}(\alpha^3)$ and lowest order differential cross
section as a function of $M_T(l\nu)$ with and without lepton
identification requirements taken into account is shown in
Fig.~\ref{fig:onea}. 
Since neutrinos do not interact with
photons, the shift in $M_W$ due to final state photon radiation is about
a factor two smaller than in the Z case. For more details we refer the reader
to Ref.~\cite{Baur:1999kt}.

\section{Non-resonant Weak Corrections to W Production and W width
Measurement}

The virtual radiative corrections to
$p\,p\hskip-7pt\hbox{$^{^{(\!-\!)}}$} \to W^{\pm} \to l^{\pm} \nu$
comprise resonant and non-resonant contributions. The non-resonant
contributions, which include those originating from the $WZ$ box
diagrams, vanish for $\hat s=M_W^2$, and thus were not included in
Ref.~\cite{Baur:1999kt}. For precision physics away from the
W pole, however, the non-resonant corrections must be included.  These
corrections  
become important at large values of $M_T$ due to the presence of large
Sudakov-like electroweak logarithms of the form $\log^2(M_T/M_W)$,
which eventually may be resummed~\cite{resummation}. 

\begin{figure}[h!]
\begin{center}
\epsfig{file=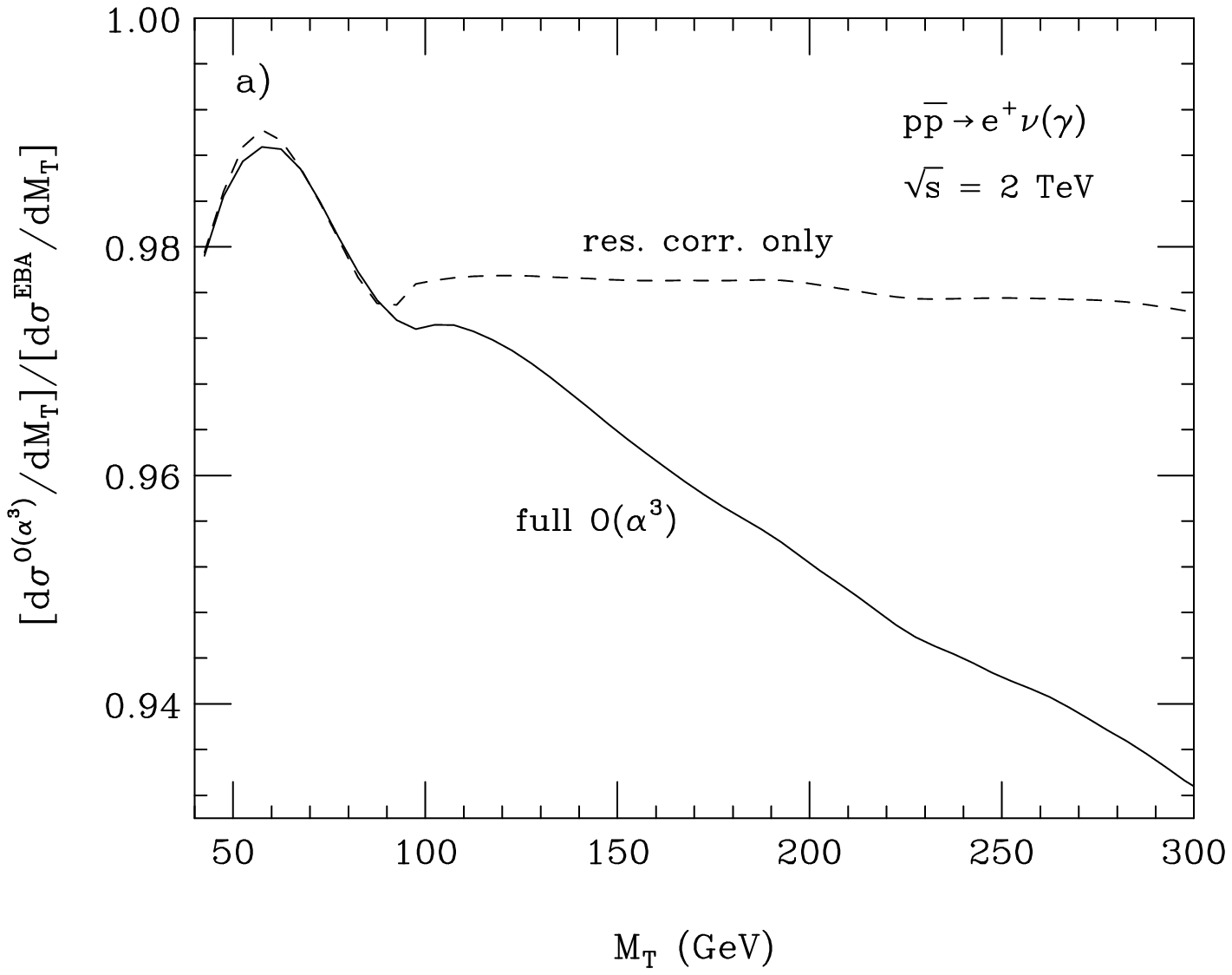,height=2.25in}\\
\epsfig{file=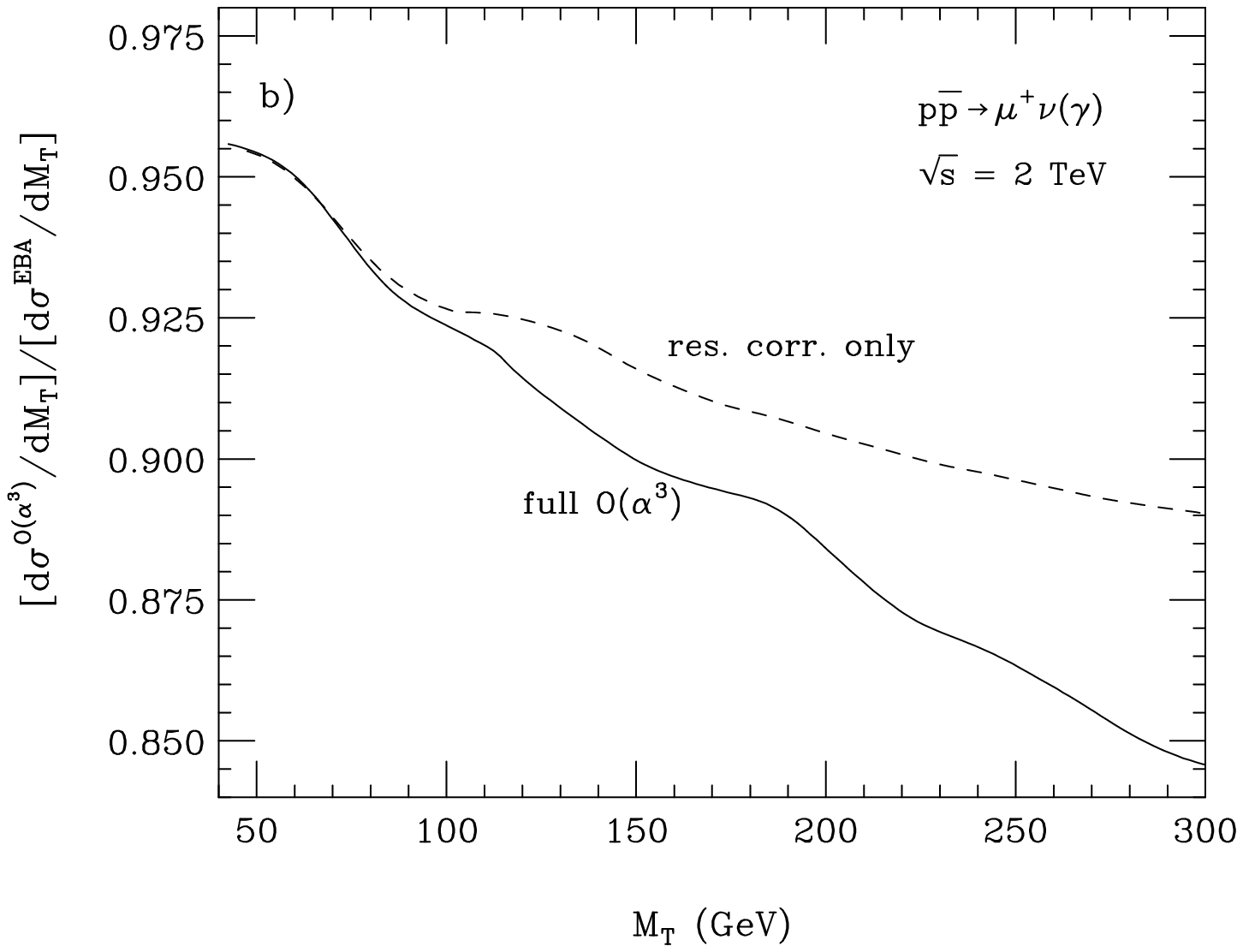,height=2.25in}
\caption[0]{The relative corrections to the transverse mass distribution
at the Tevatron a) for electrons and b) for muons in the final
state. The solid (dashed) line shows the result when the full (resonant)
${\cal O}(\alpha)$ electroweak corrections are taken into account.
 }\label{fig:two}
\end{center}
\end{figure}

In Fig.~\ref{fig:two} we show the $M_T$ distribution including the full
${\cal O}(\alpha)$ corrections 
normalized to the differential cross section in the effective Born
approximation (EBA) at the Tevatron for electron and muon final states. 
Separation
cuts and lepton identification requirements to simulate the detector
acceptance as described in Ref.~\cite{Baur:1999kt} are taken into account in 
Fig.~\ref{fig:two}. For comparison the relative
corrections including the resonant ${\cal O}(\alpha)$ electroweak
corrections only are also shown (dashed line). As expected from the
presence of large 
electroweak Sudakov-like logarithms, the non-resonant weak corrections strongly
increase in magnitude with increasing transverse masses. Both, the resonant
and non-resonant electroweak corrections reduce the differential cross 
section.

\begin{figure}[h!]
\begin{center}
\epsfig{file=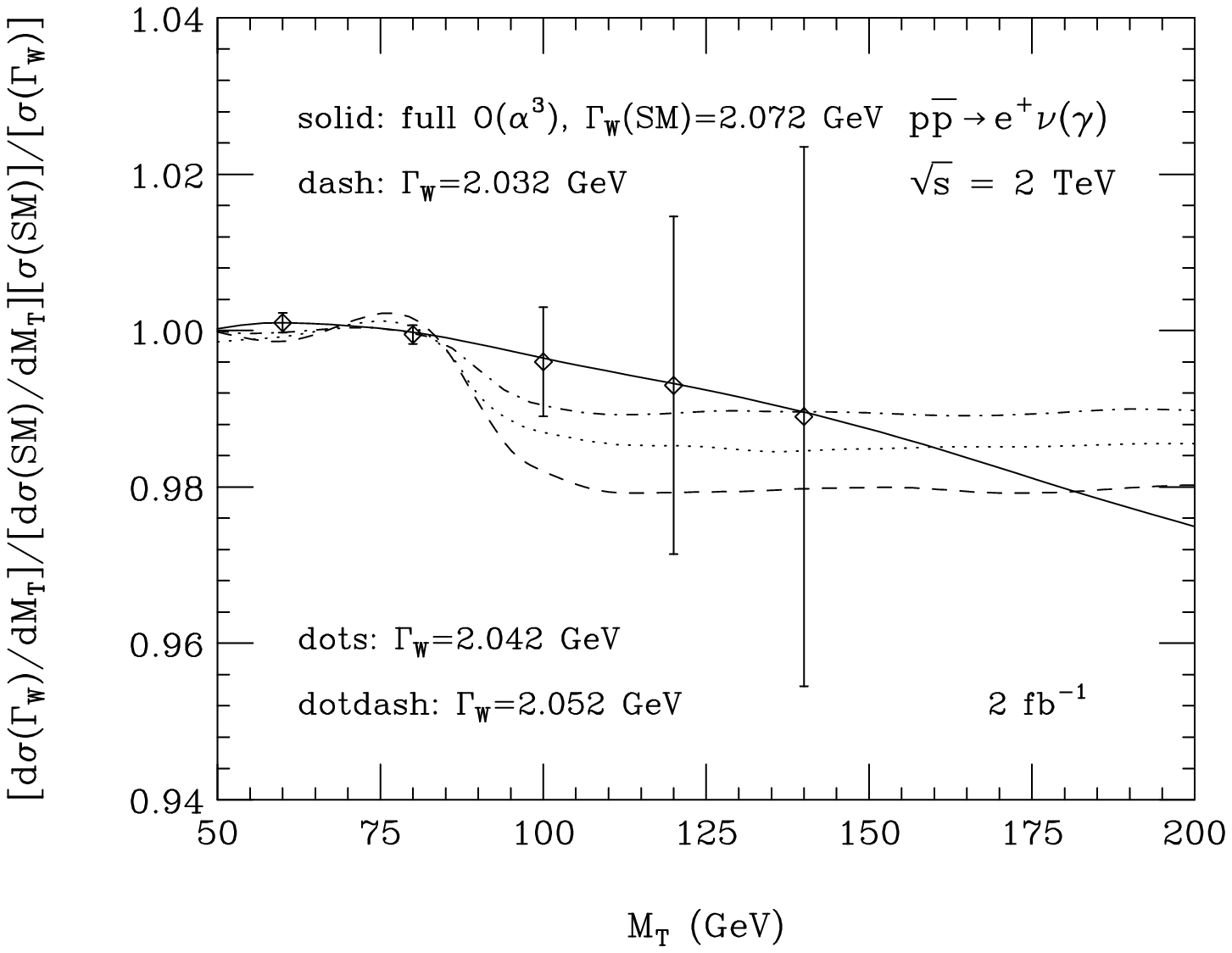,height=2.25in}
\caption[0]{The ratio of the normalized transverse mass distribution to
the normalized $M_T$ distribution with SM W width ($\Gamma_W(SM)=2.072$~GeV)
for $p\bar p\to e^+\nu_e(\gamma)$ at the Tevatron when resonant 
${\cal O}(\alpha)$ corrections only are taken into account. Similar 
results are obtained for muons in the final state. 
 }\label{fig:three}
\end{center}
\end{figure}

The high transverse mass tail can be used for a direct measurement of
the width of the W boson, $\Gamma_W$, which in the SM is predicted to be
$\Gamma_W(SM)=2.072$~GeV. In Run~I, $\Gamma_W$ has been been
measured by CDF and D\O\ with a combined uncertainty of 105~MeV from the
$M_T$ tail~\cite{comb}. In Run~IIa, with an integrated luminosity of
2~fb$^{-1}$, one expects to achieve a precision of 50~MeV
per lepton channel and experiment~\cite{Brock:1999ep}. In order to
determine how non-resonant electroweak corrections affect the
W width extracted from the tail of the transverse mass distribution, it
is useful to consider the ratio of the normalized $M_T$ distribution for
an arbitrary W width to
the normalized $M_T$ distribution with SM W width,
\begin{displaymath}
{\cal R}(\Gamma_W)={\left[d\sigma(\Gamma_W)/dM_T\right]\over
\left[d\sigma(SM)/dM_T\right]}~{\sigma(SM) \over\sigma(\Gamma_W)}
\end{displaymath}
when resonant ${\cal O}(\alpha)$ corrections only are taken into account.
It is easy to show that ${\cal R}(\Gamma_W)$ is proportional to
$\Gamma_W/\Gamma_W(SM)$ for $M_T\gg M_W$. ${\cal R}(\Gamma_W)$
is shown  for $p\bar p\to e^+\nu_e(\gamma)$ at the Tevatron in
Fig.~\ref{fig:three}.  
The dashed, dotted and dash-dotted curves represent the predictions for
$\Gamma_W=2.032$~GeV, 2.042~GeV and 2.052~GeV, respectively. 
The solid line shows the result if, instead of a non-standard W width
and resonant corrections, 
the full ${\cal O}(\alpha)$ electroweak corrections with
$\Gamma_W(SM)=2.072$~GeV are used. The error bars
represent the expected statistical uncertainties for 2~fb$^{-1}$. 
Experimental lepton
identification criteria and detector resolution effects are taken into
account in Fig.~\ref{fig:three}. 

Figure~\ref{fig:three} shows that both, non-resonant electroweak
corrections and a W width smaller than that predicted in the SM, lead to
similar effects in the tail of the $M_T$ distribution. This effect
sets on gradually in case of the non-resonant corrections. For a
non-standard W width in contrast, the ratio drops rapidly above the W
pole position, 
and becomes constant for $M_T>110$~GeV. Since the experimental precision
in the tail of the $M_T$ distribution quickly decreases with increasing
values of the transverse mass, one concludes that non-resonant
electroweak corrections, if ignored in the experimental determination of
the W width from the high $M_T$ tail, would shift the extracted value 
of $\Gamma_W$ by about $-10$~MeV. 

\section{Conclusions}

Calculations of the full ${\cal O}(\alpha)$ electroweak corrections
to W and Z boson production in hadronic collisions are now available.
Our results show that, for future precision measurements the full
electroweak ${\cal O}(\alpha)$ corrections and probably also multiple photon
radiation effects should be taken into account.  The calculation of the
non-resonant ${\cal O}(\alpha)$ corrections to W production has recently
been completed~\cite{Baurinprep} (see also Ref.~\cite{Dittmaier:2001ay}).  
As a first step towards a calculation of 
the ${\cal O}(\alpha^2)$ QED corrections, the effects of
two-photon radiation in W and Z boson production at
hadron colliders have been computed in Ref.~\cite{Baur:2000hm}.

\section{Acknowledgments}
We would like to thank the organizers of RADCOR-2002, 
for a delightful and inspiring conference experience. One of us (U.B.) is 
grateful to the Fermilab Theory Group,
where part of this work was carried out, for its generous hospitality
and for financial support. This research was supported in part by the
National Science Foundation under grant No. PHY9970703.

\end{document}